\documentclass[twocolumn,aps,prl,showpacs,superscriptaddress]{revtex4}

\usepackage{graphicx}
\usepackage{SIunits}
\usepackage{amsmath}
\usepackage{color}

\begin{document}

\title{Electromagnetically Induced Transparency and Absorption in Metamaterials:\\
       The Radiating Two-Oscillator Model and Its Experimental Confirmation}

\author{Philippe Tassin}
\email{tassin@ameslab.gov}
\affiliation{Ames Laboratory--U.S. DOE and Department of Physics and Astronomy, Iowa State University, Ames, IA 50011, USA}
\author{Lei Zhang}
\affiliation{Ames Laboratory--U.S. DOE and Department of Physics and Astronomy, Iowa State University, Ames, IA 50011, USA}
\author{Rongkuo Zhao}
\altaffiliation[Present address: ]{The Blackett Laboratory, Department of Physics, Imperial College London, London SW7 2AZ, UK}
\affiliation{Ames Laboratory--U.S. DOE and Department of Physics and Astronomy, Iowa State University, Ames, IA 50011, USA}
\author{Aditya Jain}
\affiliation{Ames Laboratory--U.S. DOE and Department of Physics and Astronomy, Iowa State University, Ames, IA 50011, USA}
\author{Thomas Koschny}
\affiliation{Ames Laboratory--U.S. DOE and Department of Physics and Astronomy, Iowa State University, Ames, IA 50011, USA}
\author{Costas M. Soukoulis}
\affiliation{Ames Laboratory--U.S. DOE and Department of Physics and Astronomy, Iowa State University, Ames, IA 50011, USA}
\affiliation{Institute of Electronic Structure and Lasers (IESL), FORTH,
             71110~Heraklion, Crete, Greece}

\date{\today}

\begin{abstract}

Several classical analogues of electromagnetically induced transparency in metamaterials have been demonstrated. A simple two-resonator model can describe their absorption spectrum qualitatively, but fails to provide information about the scattering properties---e.g., transmission and group delay. Here we develop an alternative model that rigorously includes the coupling of the radiative resonator to the external electromagnetic fields. This radiating two-oscillator model can describe both the absorption spectrum and the scattering parameters quantitatively. The model also predicts metamaterials with a narrow spectral feature in the absorption larger than the background absorption of the radiative element. This classical analogue of electromagnetically induced absorption is shown to occur when both the dissipative loss of the radiative resonator and the coupling strength are small. These predictions are subsequently demonstrated in experiments.

\end{abstract}


\maketitle

Electromagnetically induced transparency (EIT) is an effect that renders an otherwise opaque medium transparent in a narrow transmission window with low absorption and steep dispersion~\cite{Harris-1997}. It has attracted quite some interest, because of its promise for a low-loss slow-light medium. EIT was first demonstrated in certain three-level atomic systems like alkali vapors, where destructive interference between two radiative transitions creates a dark dressed superposition state with no electric dipole moment~\cite{Harris-1990,Mandel-2001,Matsko-2001,Fleischhauer-2005}. Quantum-mechanical EIT allows for the slowdown of light to a group velocity of about \unit{17}{\meter\per\second}~\cite{Hau-1999} and even for the storage of light~\cite{Fleischhauer-2000,Liu-2001,Bajcsy-2003}, but it requires complicated experimental handling because of the rather short coherence times of the superposition state.

However, it was realized soon that the characteristic features---simultaneously low absorption and steep dispersion---can also be realized in classical systems such as coupled mechanical or electrical resonators~\cite{GarridoAlzar-2002} or even coupled acoustic resonators~\cite{LiuFengming-2010}. This has led to the demonstration of many classical analogues of EIT, e.g., in electromagnetic metamaterials~\cite{Papasimakis-2008,Tassin-2009,Tassin-2009b,Singh-2009,Liu-2009,Liu-2010,Zhang-2010,Tsakmakidis-2010,Bai-2010,ZhangJingjing-2010,Kurter-2011,Kazufumi-2011} and optical microresonators~\cite{Xu-2006,Zhang-2008,Weis-2010,Eichenfield-2010,Agarwal-2010,Kekatpure-2010,Verellen-2011}. A simple model to describe these systems is a set of two coupled harmonic oscillators,
\begin{align}
\omega_{\textrm{r}}^{-2} \ddot{p}(t) + \gamma_\textrm{r}\omega_{\textrm{r}}^{-1} \dot{p}(t) + p(t) &= f(t) -\kappa q(t),\label{Eq:p}\\
\omega_{\textrm{d}}^{-2} \ddot{q}(t) + \gamma_\textrm{d}\omega_{\textrm{d}}^{-1} \dot{q}(t) + q(t) &= -\kappa p(t).\label{Eq:q}
\end{align}
The radiative resonator with resonance frequency $\omega_{\textrm{r}}$ and damping factor $\gamma_\textrm{r}$ is described by the excitation $p(t)$ and is driven by the external force $f(t)$. The dark resonator with resonance frequency $\omega_{\textrm{d}}$ and damping factor $\gamma_\textrm{d}$ is described by the excitation $q(t)$. Both resonators are linearly coupled with coupling strength $\kappa$. The individual oscillators can be mechanical, molecular, or subwavelength electromagnetic elements. The excitations would then represent the corresponding physical quantities such as the displacement from the rest position (mechanical), or the microscopic electric or magnetic dipole moment. This model reflects the essential ingredients of EIT: two coupled resonances that are are asymmetrically driven by the external force. Eqs.~(\ref{Eq:p}) and (\ref{Eq:q}) can be solved in the frequency domain by assuming a solution of the form $p(t)=\tilde{p}(\omega) \exp(-\mathrm{i}\omega t)$ and $q(t)=\tilde{q}(\omega) \exp(-\mathrm{i}\omega t)$:
\begin{equation}
\begin{split}
\tilde{p}(\omega) &= \frac{D_\mathrm{d}(\omega) \tilde{f}(\omega)}{D_\mathrm{d}(\omega) D_\mathrm{r}(\omega) - \kappa^2},\\
\tilde{q}(\omega) &= \frac{\kappa \tilde{f}(\omega)}{D_\mathrm{d}(\omega) D_\mathrm{r}(\omega) - \kappa^2},
\end{split}
\end{equation}
where $D_\mathrm{r}(\omega) = 1 - (\omega/\omega_\mathrm{r})^2 - \mathrm{i}\gamma_\mathrm{r}(\omega/\omega_\mathrm{r})$ and $D_\mathrm{d} = 1 - (\omega/\omega_\mathrm{d})^2 - \mathrm{i}\gamma_\mathrm{d}(\omega/\omega_\mathrm{d})$. The dissipated power per unit cell, which can be obtained from $Q = \tfrac{\omega^2}{2}(\gamma_\mathrm{r}|\tilde{p}(\omega)|^2+\gamma_\mathrm{d}|\tilde{q}(\omega)|^2)$, has a Lorentzian shape with a sharp incision at the resonance frequency if $\omega_{\textrm{r}} \approx \omega_{\textrm{d}}$, $\gamma_\textrm{d} \ll \gamma_\textrm{r}$, and $\gamma_\textrm{d}\gamma_\textrm{r} \ll \kappa^2 \ll 1$.

Even though the two-resonator model can qualitatively describe the absorption of classical EIT analogues, it fails to model scattering parameters of metamaterials exhibiting a classical EIT response. This is especially troublesome since it makes it impossible to determine the group delay, quite an essential parameter for slow-light media.

In this Letter, we will develop a slightly more complex model for classical EIT media. In contrast to the microscopic two-oscillator model, which does not contain information about the actual coupling to the external world, we introduce our new \emph{radiating two-oscillator model} that rigorously describes both the microscopic and the macroscopic response in terms of the radiated field (i.e., the incident, reflected, and transmitted waves). We start its derivation by recognizing that most of the EIT metamaterials fabricated to date are essentially single-layer structures rather than bulk media. Hence, their effective response can better be described by an electric current sheet with surface conductivity $\sigma_\textrm{se}$ (we restrict the discussion here to metamaterials with an electric dipole response, but the same analysis can be applied to metamaterials with a magnetic dipole response). The scattering parameters of an electric current sheet are~\cite{Tassin-2012}
\begin{equation}
R = -\frac{\zeta\sigma_\textrm{se}}{2 + \zeta\sigma_\textrm{se}}, \;
T = \frac{2}{2 + \zeta\sigma_\textrm{se}}\label{Eq:RT},
\end{equation}
where $\zeta$ is the wave impedance of the external waves. Equations~(\ref{Eq:RT}) serve as the \emph{world model}; i.e., they describe the interaction of the medium with the external electromagnetic field. The microscopic behavior of the EIT medium can still be described by the two-resonator model as given by Eqs.~(\ref{Eq:p}) and (\ref{Eq:q}).

\begin{figure*}
\centering
\includegraphics[clip]{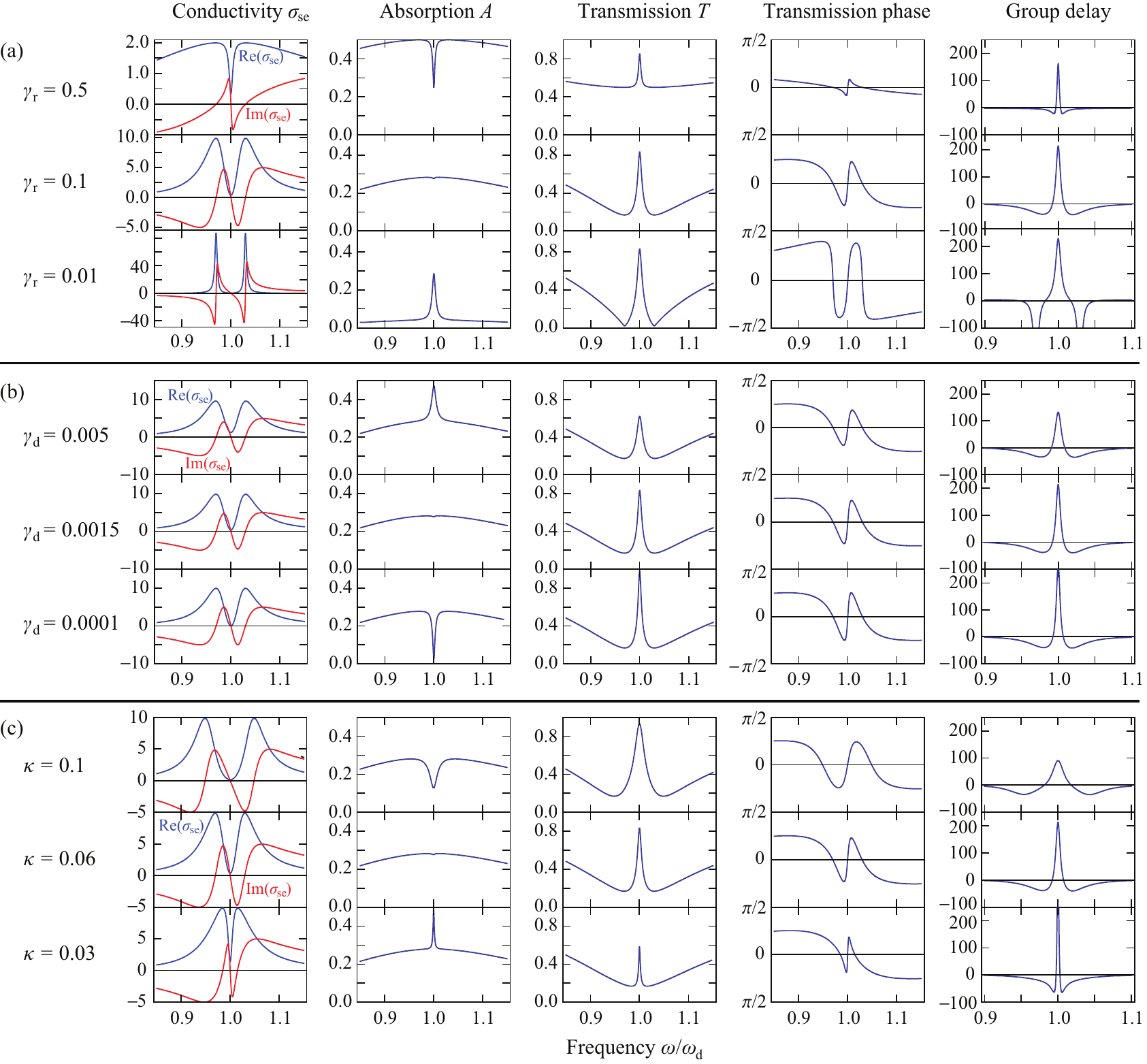}
\caption{(Color online) Spectra of the surface conductivity, absorption, transmission amplitude and phase, and group delay of EIT or EIA metamaterials as described by the radiating two-oscillator model. (a)~As a function of the dissipative damping factor of the \textit{radiative} resonator. ($\gamma_\mathrm{d} = 0.0015$, $\kappa = 0.06$, and $\beta = 1$.) (b)~As a function of the dissipative damping factor of the \textit{dark} resonator. ($\gamma_\mathrm{r} = 0.1$, $\kappa = 0.06$, and $\beta = 1$.) (c)~As a function of the coupling strength. ($\gamma_\mathrm{d} = 0.0015$, $\gamma_\mathrm{r} = 0.1$, and $\beta = 1$.)}\label{Fig:ParameterDependence}
\end{figure*}

In order to complete the radiating two-oscillator model, we have to find a connection between the external behavior of the system (the surface field $E_\mathrm{s}$ and the surface conductivity $\sigma_\textrm{se}$) and the microscopic behavior (the excitations $p$ and $q$ and the driving force $f$). The macroscopic surface field $E_\mathrm{s}$ is the spatially averaged electric field on the current sheet---it is related to the incident field by $E_\mathrm{s} = T \times E_{in}$ and can be calculated in this way both for experiments and simulations. First, we observe that each of the constituent meta-atoms contributes a dipole moment $p$ to the metamaterial and, if there are $n_\mathrm{s}$ atoms per unit of surface area, the average polarization current thus equals
\begin{equation}
\left<j_\mathrm{s}(t)\right> = n_\mathrm{s}\dot{p}(t) \leftrightarrow \left<\tilde{j}_\mathrm{s}(\omega)\right> = -\mathrm{i}\omega n_\mathrm{s}\tilde{p}(\omega).\label{Eq:SurfaceCurrent}
\end{equation}
The dark resonator does not contribute to the surface current since it has no dipole moment commensurate with the external field. Secondly, we need to find a connection between the surface field $E_\mathrm{s}$, which drives the dipole oscillation in the world model, and the driving force $f$ in the microscopic model; i.e., we seek the proportionality constant $C$ in $f(t) = C E_\mathrm{s}(t)$. (Note that the surface field $E_\mathrm{s}$ is different from the incident field because of the scattering from the meta-atoms.) This can be done by recalling that for our linear meta-atom the average dipole moment must be proportional to the electric field at the surface: $n_\mathrm{s}\tilde{p}(\omega) = \epsilon_0\chi_\mathrm{se}(\omega)\tilde{E}_\mathrm{s}(\omega)$, where $\chi_\mathrm{se}$ is the surface susceptibility. In the static limit, this yields
\begin{equation}
\epsilon_0\chi_\mathrm{se}^\mathrm{(static)}\tilde{E_\mathrm{s}}(0) = n_\mathrm{s}\tilde{p}(0) = n_\mathrm{s}\left( 1 - \kappa^2 \right)^{-1} \tilde{f}(0) \approx n_\mathrm{s} \tilde{f}(0),\label{Eq:SurfacePolarisation}
\end{equation}
where we used the fact that $\kappa \ll 1$ under EIT conditions in the last approximation. Using Eqs.~(\ref{Eq:SurfaceCurrent}) and (\ref{Eq:SurfacePolarisation}), we can now determine the surface conductivity from the constitutive equation $\left<\tilde{j}_\mathrm{s}(\omega)\right> = \sigma_\mathrm{se}\tilde{E}_\mathrm{s}(\omega)$:
\begin{equation}
\sigma_\mathrm{se} \approx \epsilon_0\chi_\mathrm{se}^\mathrm{(static)}\frac{-\mathrm{i}\omega\tilde{p}(\omega)}{\tilde{f}(\omega)}
                   = \frac{-\mathrm{i}\omega \beta D_\mathrm{d}(\omega)}{D_\mathrm{d}(\omega) D_\mathrm{r}(\omega) - \kappa^2},
\end{equation}
where $\beta \equiv \epsilon_0\chi_\mathrm{se}^\mathrm{(static)}$. Once we have determined the surface conductivity, we can calculate the scattering parameters from Eqs.~(\ref{Eq:RT}) and other derived quantities, such as the absorbance and the group delay:
\begin{align}
A &= 1 - \left|T\right|^2 - \left|R\right|^2 = \left|T\right|^2 \mathrm{Re}\left(\zeta\sigma_\mathrm{se}\right),\\
\tau_\mathrm{g} &= \mathrm{Im}\left(\frac{\mathrm{d} \ln T}{\mathrm{d}\omega}\right) = -\frac{1}{2}\mathrm{Im}\left(T \frac{\mathrm{d} \zeta\sigma_\mathrm{se}}{\mathrm{d}\omega}\right).
\end{align}

Note that, from the perspective of the microscopic description in terms of two coupled oscillators used in previous literature, our model introduces a radiation damping term in the bright oscillator as well as an excitation-dependent external driving force, both of which originate in the scattered field of the bright resonators responsible for the reflectance (and nonunity transmittance) of the macroscopic sample. However, even with those corrections to the two-resonator model, we still need the full radiating two-oscillator model as described in this Letter to calculate the transmittance and group delay---the two most important characteristics of an EIT system.

The radiating two-oscillator model allows us to understand the response of EIT metamaterials as a function of their microscopic parameters. In Fig.~\ref{Fig:ParameterDependence}(a), we plot the surface conductivity, the absorption, the transmission amplitude and phase, and the group delay for a set of metamaterials with different dissipative damping in the radiative resonator. In the top row (high damping), we recognize the typical features of EIT. The conductivity has a Lorentzian envelope with a sharp incision, resulting in a frequency window with large transmission and small absorption. At the same time, there is large normal dispersion in the transmission phase, which leads to a significantly enhanced group delay. The reduced response can be understood from the destructive interference of the excitation $p$ due to the external field and due to the coupling from the dark resonator (see $\left|\tilde{p}(\omega)\right|$ at the resonance frequency in Fig.~S1~\cite{SupplMat}).

An interesting phenomenon occurs if we decrease the dissipative loss factor $\gamma_\mathrm{r}$ of the radiative resonator [second row in Fig.~\ref{Fig:ParameterDependence}(a)]. There is still a frequency window with high transmission, but the incision in the absorption spectrum becomes smaller and finally disappears. This does not mean, however, that the dark resonance has disappeared, as we can clearly see from the strong dispersion and associated enhanced group delay. Rather, the background absorption of the radiative resonance is decreased, but the radiative resonance is still sufficiently broadened by the radiation damping, while the excitation of the dark resonance is barely changed. At a certain point, the absorption reduction in the radiative resonator is exactly cancelled by the absorption in the dark resonator. When we further decrease the dissipative loss of the radiative resonator, the absorption spectrum turns into a very dim background with a narrow peak at the resonance frequency $\omega = \omega_\mathrm{d}$. This phenomenon can be seen as a classical analogue of \textit{electromagnetically induced absorption} (EIA)~\cite{Akulshin-1998} and we believe it may be interesting for applications in spectroscopy and sensing since the width of the peak is reduced by the lack of radiation damping in the dark resonator and the additional narrowing due to the coupling. Note that the EIA effect could only be described by the radiating two-oscillator model, since the bare two-resonator model lacks radiative broadening of the radiative resonator.  

\begin{figure*}
\centering
\includegraphics[clip]{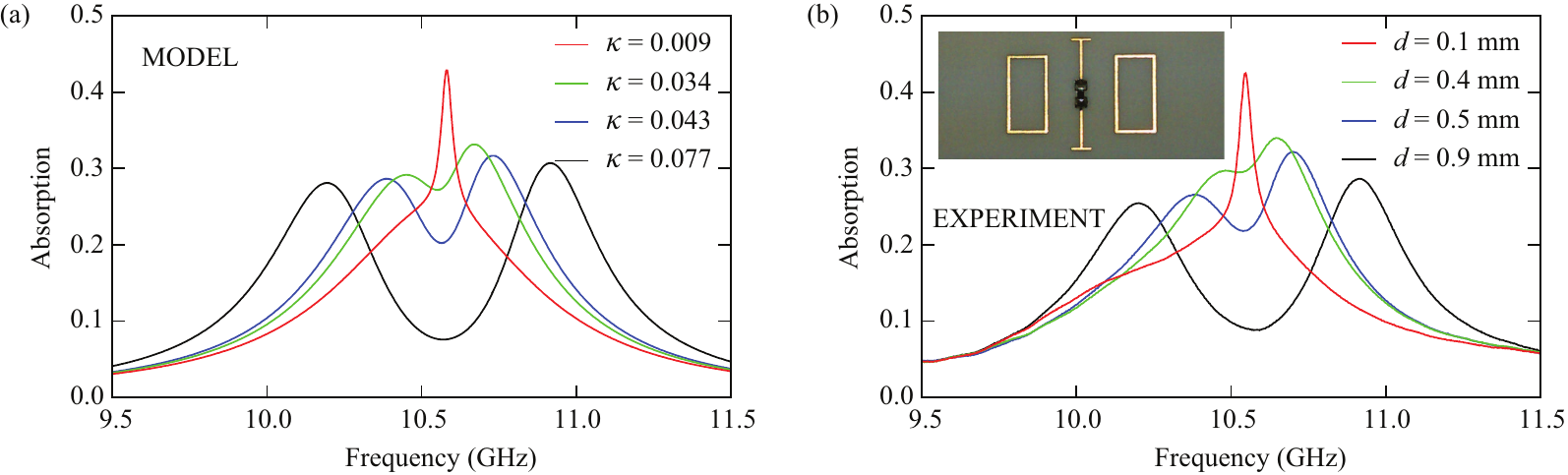}
\caption{(Color online) Absorption spectra of the EIT or EIA metamaterials for different coupling strengths. (a)~Predicted by the radiating two-oscillator model with parameters $\omega_\textrm{d} = \unit{2\pi\times 10.58}{\giga\hertz}$, $\gamma_\textrm{d} = 0.0035$, $\omega_\textrm{r} = \unit{2\pi\times 10.12}{\giga\hertz}$, $\gamma_\textrm{r} = 0.01$, and $\beta = 0.33/(\omega_\textrm{r}\eta_0)$. (b)~Measured for microwave metamaterials in the X band.}\label{Fig:ExpCoupling}
\end{figure*}

\begin{figure*}
\centering
\includegraphics[clip]{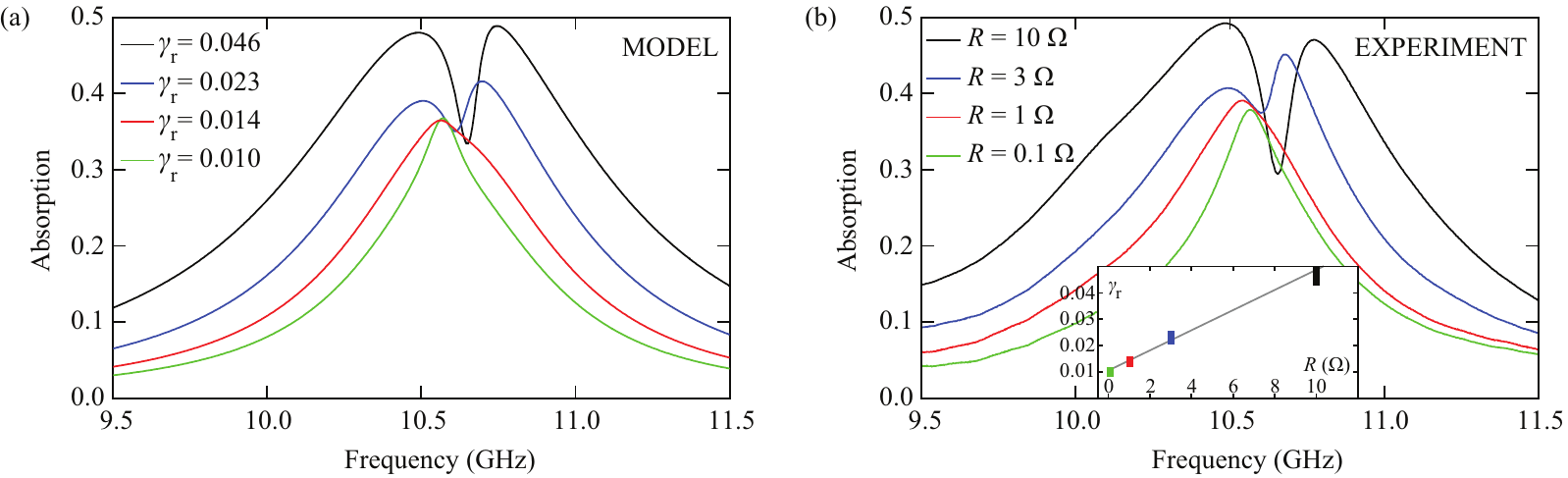}
\caption{(Color online) Absorption spectra of the EIT or EIA metamaterials for different damping of the radiative resonator. (a)~Predicted by the radiating two-oscillator model with parameters $\omega_\textrm{r} = \unit{2\pi\times 10.11}{\giga\hertz}$, $\gamma_\textrm{d} = 0.0032$, $\kappa = 0.027$, and $\beta = 0.38/(\omega_\textrm{r}\eta_0)$. (b)~Measured for microwave metamaterials in the X band. The inset shows the linear relationship between the damping constants in the model and the lumped resistances in the experiment. The error bars represent the 5\% tolerance of the resistors.}\label{Fig:ExpResistors}
\end{figure*}

The transition between EIT and EIA can also be observed when we increase the dissipative loss of the dark resonator [see Fig.~\ref{Fig:ParameterDependence}(b)]. When $\gamma_\mathrm{d}$ is increased, the radiative resonance remains unaltered, but the absorption in the center of the transparency window goes up. Eventually, the loss in the dark resonator overcomes the loss reduction due to the destructive interference in the radiative resonator. Note, however, that too large a value of $\gamma_\mathrm{d}$ destroys the EIT or EIA phenomenon as shown in Fig.~S2~\cite{SupplMat}. Finally, EIA can also be achieved by decreasing the coupling strength $\kappa$, as in Fig.~\ref{Fig:ParameterDependence}(c). Weaker coupling creates a narrower transparency window with larger excitation $q$ in the dark resonator. This in turn increases the absorption at the resonance frequency, resulting in EIA when the dissipation in the dark resonance overcomes the loss reduction in the radiative resonance. Again, EIT or EIA is destroyed when $\kappa^2 < \gamma_\mathrm{d}\gamma_\mathrm{r}$ (see Fig.~S3~\cite{SupplMat}).

Subsequently, we have confirmed the predictions of the radiating two-oscillator model in microwave metamaterials consisting of a copper cut wire as the radiative resonator and two copper closed-ring resonators as the dark resonator on a Rogers substrate~\cite{SupplMat} (see inset in Fig.~\ref{Fig:ExpCoupling}). The closed rings provide a dark resonance by using the antisymmetric hybridization of the electric dipole modes of the split-ring resonators---this hybridization is actually an electric quadrupole resonance with zero overlap with the fundamental waveguide mode (see the Supplemental Material for field plots demonstrating the dark and bright modes~\cite{SupplMat}). The measurements were performed in a WR-90 waveguide and the scattering parameters were measured using a vector network analyzer (HP E8364) and calibrated using a transmission-reflection-match (TRM) method. In Fig.~\ref{Fig:ExpCoupling}(b), we have plotted the experimental absorption spectrum as a function of the coupling strength (the coupling strength was varied by moving the cut wire over a distance $d$ horizontally between the closed rings). We see that the absorption spectrum of the model [Fig.~\ref{Fig:ExpCoupling}(a)] is in excellent agreement with the experiments [Fig.~\ref{Fig:ExpCoupling}(b)]. Figure~S5 shows that also the transmission and reflection coefficients are in excellent agreement~\cite{SupplMat}. The values of $\kappa$ from the model are perfectly proportional to the offset of the wire from the center ($d$), further confirming that the model can quantitatively describe EIT or EIA metamaterials. In a second set of experiments, we have altered the dissipative loss of the radiative resonator (by soldering SMD resistors with different resistance values $R$ into the cut wire). We again see that the experiments [Fig.~\ref{Fig:ExpResistors}(b)] convincingly reproduce the transition from EIT to EIA, and are in very good agreement with the radiating two-oscillator model [Fig.~\ref{Fig:ExpResistors}(a)]. The transmission and reflection coefficients from the model are also in excellent agreement with the experiments (see Fig.~S6~\cite{SupplMat}). The matching damping constants are plotted as a function of the resistor values in the inset of Fig.~\ref{Fig:ExpResistors}, revealing a linear relationship, though with an offset this time. The offset is because part of the dissipation in the radiative resonator happens in the capacitor due to relaxation loss in the substrate. The equivalent resistance of the relaxation loss is estimated to be \unit{3.4}{\ohm}. The experimental group delay curves are provided in Fig.~S7~\cite{SupplMat}.

In this Letter, we have focused on metamaterials with subwavelength constituents. Our model can not only quantitatively describe EIT metamaterials, but it also reveals a classical analogue of EIA---which is characterized by a sharp absorption peak on a shallow background---when the radiative resonator has small dissipative loss, but is still sufficiently broadened by radiation damping. There have recently been two papers in which a phenomenon similar to EIA is discovered when the two resonators are coupled to the external wave with different phase~\cite{Verslegers-2012} or when a retardation-induced phase shift occurs in the coupling mechanism~\cite{Taubert-2012}. We believe that our model can also describe these experiments with small changes (e.g., complex $\kappa$). Nevertheless, in truly homogenizable metamaterials, such phase differences are not possible and we must revert to radiation-broadened resonators to achieve EIA.

Work at Ames Laboratory was supported by the U.S.\ Department of Energy (Basic Energy Science, Division of Materials Sciences and Engineering) under Contract No.\ DE-AC02-07CH11358 (theoretical and computational studies) and by ONR Award No.\ N00014-10-1-0925 (experiments).


\begin{thebibliography}{33}
\expandafter\ifx\csname natexlab\endcsname\relax\def\natexlab#1{#1}\fi
\expandafter\ifx\csname bibnamefont\endcsname\relax
  \def\bibnamefont#1{#1}\fi
\expandafter\ifx\csname bibfnamefont\endcsname\relax
  \def\bibfnamefont#1{#1}\fi
\expandafter\ifx\csname citenamefont\endcsname\relax
  \def\citenamefont#1{#1}\fi
\expandafter\ifx\csname url\endcsname\relax
  \def\url#1{\texttt{#1}}\fi
\expandafter\ifx\csname urlprefix\endcsname\relax\def\urlprefix{URL }\fi
\providecommand{\bibinfo}[2]{#2}
\providecommand{\eprint}[2][]{\url{#2}}

\bibitem[{\citenamefont{{Harris}}(1997)}]{Harris-1997}
\bibinfo{author}{\bibfnamefont{S.~E.} \bibnamefont{{Harris}}},
  \bibinfo{journal}{Phys.\ Today} \textbf{\bibinfo{volume}{50}},
  \bibinfo{pages}{36} (\bibinfo{year}{1997}).

\bibitem[{\citenamefont{{Harris} et~al.}(1990)\citenamefont{{Harris}, {Field},
  and {Imamoglu}}}]{Harris-1990}
\bibinfo{author}{\bibfnamefont{S.~E.} \bibnamefont{{Harris}}},
  \bibinfo{author}{\bibfnamefont{J.~E.} \bibnamefont{{Field}}},
  \bibnamefont{and}
  \bibinfo{author}{\bibfnamefont{A.}~\bibnamefont{{Imamoglu}}},
  \bibinfo{journal}{\prl} \textbf{\bibinfo{volume}{64}}, \bibinfo{pages}{1107}
  (\bibinfo{year}{1990}).

\bibitem[{\citenamefont{Mandel}(2001)}]{Mandel-2001}
\bibinfo{author}{\bibfnamefont{P.}~\bibnamefont{Mandel}},
  \bibinfo{journal}{Hyperfine Interact.} \textbf{\bibinfo{volume}{135}},
  \bibinfo{pages}{223} (\bibinfo{year}{2001}).

\bibitem[{\citenamefont{Matsko et~al.}(2001)\citenamefont{Matsko,
  Kocharovskaya, Rostovtsev, Welch, Zibrov, and Scully}}]{Matsko-2001}
\bibinfo{author}{\bibfnamefont{A.~B.} \bibnamefont{Matsko}},
  \bibinfo{author}{\bibfnamefont{O.}~\bibnamefont{Kocharovskaya}},
  \bibinfo{author}{\bibfnamefont{Y.}~\bibnamefont{Rostovtsev}},
  \bibinfo{author}{\bibfnamefont{G.~R.} \bibnamefont{Welch}},
  \bibinfo{author}{\bibfnamefont{A.~S.} \bibnamefont{Zibrov}},
  \bibnamefont{and} \bibinfo{author}{\bibfnamefont{M.~O.}
  \bibnamefont{Scully}}, \bibinfo{journal}{Adv.\ Atom.\ Mol.\ Opt.\ Phys}
  \textbf{\bibinfo{volume}{46}}, \bibinfo{pages}{191} (\bibinfo{year}{2001}).

\bibitem[{\citenamefont{Fleischhauer et~al.}(2005)\citenamefont{Fleischhauer,
  Imamoglu, and Marangos}}]{Fleischhauer-2005}
\bibinfo{author}{\bibfnamefont{M.}~\bibnamefont{Fleischhauer}},
  \bibinfo{author}{\bibfnamefont{A.}~\bibnamefont{Imamoglu}}, \bibnamefont{and}
  \bibinfo{author}{\bibfnamefont{J.~P.} \bibnamefont{Marangos}},
  \bibinfo{journal}{\rmp} \textbf{\bibinfo{volume}{77}}, \bibinfo{pages}{633}
  (\bibinfo{year}{2005}).

\bibitem[{\citenamefont{{Hau} et~al.}(1999)\citenamefont{{Hau}, {Harris},
  {Dutton}, and {Behroozi}}}]{Hau-1999}
\bibinfo{author}{\bibfnamefont{L.~V.} \bibnamefont{{Hau}}},
  \bibinfo{author}{\bibfnamefont{S.~E.} \bibnamefont{{Harris}}},
  \bibinfo{author}{\bibfnamefont{Z.}~\bibnamefont{{Dutton}}}, \bibnamefont{and}
  \bibinfo{author}{\bibfnamefont{C.~H.} \bibnamefont{{Behroozi}}},
  \bibinfo{journal}{\nat} \textbf{\bibinfo{volume}{397}}, \bibinfo{pages}{594}
  (\bibinfo{year}{1999}).

\bibitem[{\citenamefont{{Fleischhauer} and {Lukin}}(2000)}]{Fleischhauer-2000}
\bibinfo{author}{\bibfnamefont{M.}~\bibnamefont{{Fleischhauer}}}
  \bibnamefont{and} \bibinfo{author}{\bibfnamefont{M.~D.}
  \bibnamefont{{Lukin}}}, \bibinfo{journal}{\prl}
  \textbf{\bibinfo{volume}{84}}, \bibinfo{pages}{5094} (\bibinfo{year}{2000}).

\bibitem[{\citenamefont{{Liu} et~al.}(2001)\citenamefont{{Liu}, {Dutton},
  {Behroozi}, and {Hau}}}]{Liu-2001}
\bibinfo{author}{\bibfnamefont{C.}~\bibnamefont{{Liu}}},
  \bibinfo{author}{\bibfnamefont{Z.}~\bibnamefont{{Dutton}}},
  \bibinfo{author}{\bibfnamefont{C.~H.} \bibnamefont{{Behroozi}}},
  \bibnamefont{and} \bibinfo{author}{\bibfnamefont{L.~V.} \bibnamefont{{Hau}}},
  \bibinfo{journal}{\nat} \textbf{\bibinfo{volume}{409}}, \bibinfo{pages}{490}
  (\bibinfo{year}{2001}).

\bibitem[{\citenamefont{Bajcsy et~al.}(2003)\citenamefont{Bajcsy, Zibrov, and
  Lukin}}]{Bajcsy-2003}
\bibinfo{author}{\bibfnamefont{M.}~\bibnamefont{Bajcsy}},
  \bibinfo{author}{\bibfnamefont{A.~S.} \bibnamefont{Zibrov}},
  \bibnamefont{and} \bibinfo{author}{\bibfnamefont{M.~D.} \bibnamefont{Lukin}},
  \bibinfo{journal}{\nat} \textbf{\bibinfo{volume}{426}}, \bibinfo{pages}{638}
  (\bibinfo{year}{2003}).

\bibitem[{\citenamefont{{Garrido Alzar} et~al.}(2002)\citenamefont{{Garrido
  Alzar}, {Martinez}, and {Nussensveig}}}]{GarridoAlzar-2002}
\bibinfo{author}{\bibfnamefont{C.~L.} \bibnamefont{{Garrido Alzar}}},
  \bibinfo{author}{\bibfnamefont{M.~A.~G.} \bibnamefont{{Martinez}}},
  \bibnamefont{and}
  \bibinfo{author}{\bibfnamefont{P.}~\bibnamefont{{Nussensveig}}},
  \bibinfo{journal}{Am.\ J.\ Phys.} \textbf{\bibinfo{volume}{70}},
  \bibinfo{pages}{37} (\bibinfo{year}{2002}).

\bibitem[{\citenamefont{Liu et~al.}(2010{\natexlab{a}})\citenamefont{Liu, Ke,
  Zhang, Wen, Shi, Liu, and Sheng}}]{LiuFengming-2010}
\bibinfo{author}{\bibfnamefont{F.}~\bibnamefont{Liu}},
  \bibinfo{author}{\bibfnamefont{M.}~\bibnamefont{Ke}},
  \bibinfo{author}{\bibfnamefont{A.}~\bibnamefont{Zhang}},
  \bibinfo{author}{\bibfnamefont{W.}~\bibnamefont{Wen}},
  \bibinfo{author}{\bibfnamefont{J.}~\bibnamefont{Shi}},
  \bibinfo{author}{\bibfnamefont{Z.}~\bibnamefont{Liu}}, \bibnamefont{and}
  \bibinfo{author}{\bibfnamefont{P.}~\bibnamefont{Sheng}},
  \bibinfo{journal}{\pre} \textbf{\bibinfo{volume}{82}},
  \bibinfo{pages}{026601} (\bibinfo{year}{2010}{\natexlab{a}}).

\bibitem[{\citenamefont{{Papasimakis} et~al.}(2008)\citenamefont{{Papasimakis},
  {Fedotov}, {Zheludev}, and {Prosvirnin}}}]{Papasimakis-2008}
\bibinfo{author}{\bibfnamefont{N.}~\bibnamefont{{Papasimakis}}},
  \bibinfo{author}{\bibfnamefont{V.~A.} \bibnamefont{{Fedotov}}},
  \bibinfo{author}{\bibfnamefont{N.~I.} \bibnamefont{{Zheludev}}},
  \bibnamefont{and} \bibinfo{author}{\bibfnamefont{S.~L.}
  \bibnamefont{{Prosvirnin}}}, \bibinfo{journal}{\prl}
  \textbf{\bibinfo{volume}{101}}, \bibinfo{pages}{253903}
  (\bibinfo{year}{2008}).

\bibitem[{\citenamefont{{Tassin}
  et~al.}(2009{\natexlab{a}})\citenamefont{{Tassin}, {Zhang}, {Koschny},
  {Economou}, and {Soukoulis}}}]{Tassin-2009}
\bibinfo{author}{\bibfnamefont{P.}~\bibnamefont{{Tassin}}},
  \bibinfo{author}{\bibfnamefont{L.}~\bibnamefont{{Zhang}}},
  \bibinfo{author}{\bibfnamefont{T.}~\bibnamefont{{Koschny}}},
  \bibinfo{author}{\bibfnamefont{E.~N.} \bibnamefont{{Economou}}},
  \bibnamefont{and} \bibinfo{author}{\bibfnamefont{C.~M.}
  \bibnamefont{{Soukoulis}}}, \bibinfo{journal}{\prl}
  \textbf{\bibinfo{volume}{102}}, \bibinfo{pages}{053901}
  (\bibinfo{year}{2009}{\natexlab{a}}).

\bibitem[{\citenamefont{{Tassin}
  et~al.}(2009{\natexlab{b}})\citenamefont{{Tassin}, {Zhang}, {Koschny},
  {Economou}, and {Soukoulis}}}]{Tassin-2009b}
\bibinfo{author}{\bibfnamefont{P.}~\bibnamefont{{Tassin}}},
  \bibinfo{author}{\bibfnamefont{L.}~\bibnamefont{{Zhang}}},
  \bibinfo{author}{\bibfnamefont{T.}~\bibnamefont{{Koschny}}},
  \bibinfo{author}{\bibfnamefont{E.~N.} \bibnamefont{{Economou}}},
  \bibnamefont{and} \bibinfo{author}{\bibfnamefont{C.~M.}
  \bibnamefont{{Soukoulis}}}, \bibinfo{journal}{Opt.\ Express}
  \textbf{\bibinfo{volume}{17}}, \bibinfo{pages}{5595}
  (\bibinfo{year}{2009}{\natexlab{b}}).

\bibitem[{\citenamefont{{Liu}
  et~al.}(2009{\natexlab{b}})\citenamefont{{Liu}, {Langguth}, {Weiss},
  {Kastel}, {Fleischhauer}, {Pfau}, and {Giessen}}}]{Liu-2009}
\bibinfo{author}{\bibfnamefont{N.}~\bibnamefont{{Liu}}},
  \bibinfo{author}{\bibfnamefont{L.}~\bibnamefont{{Langguth}}},
  \bibinfo{author}{\bibfnamefont{T.}~\bibnamefont{{Weiss}}},
  \bibinfo{author}{\bibfnamefont{J.} \bibnamefont{{Kastel}}},
  \bibinfo{author}{\bibfnamefont{T.}~\bibnamefont{{Fleischhauer}}},
  \bibinfo{author}{\bibfnamefont{J.} \bibnamefont{{Pfau}}},
  \bibnamefont{and} \bibinfo{author}{\bibfnamefont{C.~M.}
  \bibnamefont{{Giessen}}}, \bibinfo{journal}{Nature Mater.}
  \textbf{\bibinfo{volume}{8}}, \bibinfo{pages}{758}
  (\bibinfo{year}{2009}{\natexlab{b}}).

\bibitem[{\citenamefont{Liu et~al.}(2010{\natexlab{b}})\citenamefont{Liu,
  Weiss, Mesch, Langguth, Eigenthaler, Hirscher, Sonnichsen, and
  Giessen}}]{Liu-2010}
\bibinfo{author}{\bibfnamefont{N.}~\bibnamefont{Liu}},
  \bibinfo{author}{\bibfnamefont{T.}~\bibnamefont{Weiss}},
  \bibinfo{author}{\bibfnamefont{M.}~\bibnamefont{Mesch}},
  \bibinfo{author}{\bibfnamefont{L.}~\bibnamefont{Langguth}},
  \bibinfo{author}{\bibfnamefont{U.}~\bibnamefont{Eigenthaler}},
  \bibinfo{author}{\bibfnamefont{M.}~\bibnamefont{Hirscher}},
  \bibinfo{author}{\bibfnamefont{C.}~\bibnamefont{Sonnichsen}},
  \bibnamefont{and} \bibinfo{author}{\bibfnamefont{H.}~\bibnamefont{Giessen}},
  \bibinfo{journal}{Nano Lett.} \textbf{\bibinfo{volume}{10}},
  \bibinfo{pages}{1103} (\bibinfo{year}{2010}{\natexlab{b}}).

\bibitem[{\citenamefont{{Singh} et~al.}(2009)\citenamefont{{Singh},
  {Rockstuhl}, {Lederer}, and {Zhang}}}]{Singh-2009}
\bibinfo{author}{\bibfnamefont{R.}~\bibnamefont{{Singh}}},
  \bibinfo{author}{\bibfnamefont{C.}~\bibnamefont{{Rockstuhl}}},
  \bibinfo{author}{\bibfnamefont{F.}~\bibnamefont{{Lederer}}},
  \bibnamefont{and} \bibinfo{author}{\bibfnamefont{W.}~\bibnamefont{{Zhang}}},
  \bibinfo{journal}{\prb} \textbf{\bibinfo{volume}{79}},
  \bibinfo{pages}{085111} (\bibinfo{year}{2009}).

\bibitem[{\citenamefont{Zhang et~al.}(2010{\natexlab{a}})\citenamefont{Zhang,
  Tassin, Koschny, Kurter, Anlage, and Soukoulis}}]{Zhang-2010}
\bibinfo{author}{\bibfnamefont{L.}~\bibnamefont{Zhang}},
  \bibinfo{author}{\bibfnamefont{P.}~\bibnamefont{Tassin}},
  \bibinfo{author}{\bibfnamefont{T.}~\bibnamefont{Koschny}},
  \bibinfo{author}{\bibfnamefont{C.}~\bibnamefont{Kurter}},
  \bibinfo{author}{\bibfnamefont{S.~M.} \bibnamefont{Anlage}},
  \bibnamefont{and} \bibinfo{author}{\bibfnamefont{C.~M.}
  \bibnamefont{Soukoulis}}, \bibinfo{journal}{\apl}
  \textbf{\bibinfo{volume}{97}}, \bibinfo{pages}{241904}
  (\bibinfo{year}{2010}{\natexlab{a}}).

\bibitem[{\citenamefont{Tsakmakidis et~al.}(2010)\citenamefont{Tsakmakidis,
  Wartak, Cook, Hamm, and Hess}}]{Tsakmakidis-2010}
\bibinfo{author}{\bibfnamefont{K.~L.} \bibnamefont{Tsakmakidis}},
  \bibinfo{author}{\bibfnamefont{M.~S.} \bibnamefont{Wartak}},
  \bibinfo{author}{\bibfnamefont{J.~J.~H.} \bibnamefont{Cook}},
  \bibinfo{author}{\bibfnamefont{J.~M.} \bibnamefont{Hamm}}, \bibnamefont{and}
  \bibinfo{author}{\bibfnamefont{O.}~\bibnamefont{Hess}},
  \bibinfo{journal}{\prb} \textbf{\bibinfo{volume}{81}},
  \bibinfo{pages}{195128} (\bibinfo{year}{2010}).

\bibitem[{\citenamefont{Bai et~al.}(2010)\citenamefont{Bai, Liu, Chen, Cheng,
  Kang, and Wang}}]{Bai-2010}
\bibinfo{author}{\bibfnamefont{Q.}~\bibnamefont{Bai}},
  \bibinfo{author}{\bibfnamefont{C.}~\bibnamefont{Liu}},
  \bibinfo{author}{\bibfnamefont{J.}~\bibnamefont{Chen}},
  \bibinfo{author}{\bibfnamefont{C.}~\bibnamefont{Cheng}},
  \bibinfo{author}{\bibfnamefont{M.}~\bibnamefont{Kang}}, \bibnamefont{and}
  \bibinfo{author}{\bibfnamefont{H.-T.} \bibnamefont{Wang}},
  \bibinfo{journal}{J.\ Appl.\ Phys.} \textbf{\bibinfo{volume}{107}},
  \bibinfo{pages}{093104} (\bibinfo{year}{2010}).

\bibitem[{\citenamefont{Zhang et~al.}(2010{\natexlab{b}})\citenamefont{Zhang,
  Xiao, Jeppesen, Kristensen, and Mortensen}}]{ZhangJingjing-2010}
\bibinfo{author}{\bibfnamefont{J.}~\bibnamefont{Zhang}},
  \bibinfo{author}{\bibfnamefont{S.}~\bibnamefont{Xiao}},
  \bibinfo{author}{\bibfnamefont{C.}~\bibnamefont{Jeppesen}},
  \bibinfo{author}{\bibfnamefont{A.}~\bibnamefont{Kristensen}},
  \bibnamefont{and} \bibinfo{author}{\bibfnamefont{N.~A.}
  \bibnamefont{Mortensen}}, \bibinfo{journal}{Opt.\ Express}
  \textbf{\bibinfo{volume}{18}}, \bibinfo{pages}{17187}
  (\bibinfo{year}{2010}{\natexlab{b}}).

\bibitem[{\citenamefont{Kurter et~al.}(2011)\citenamefont{Kurter, Tassin,
  Zhang, Koschny, Zhuravel, Ustinov, Anlage, and Soukoulis}}]{Kurter-2011}
\bibinfo{author}{\bibfnamefont{C.}~\bibnamefont{Kurter}},
  \bibinfo{author}{\bibfnamefont{P.}~\bibnamefont{Tassin}},
  \bibinfo{author}{\bibfnamefont{L.}~\bibnamefont{Zhang}},
  \bibinfo{author}{\bibfnamefont{T.}~\bibnamefont{Koschny}},
  \bibinfo{author}{\bibfnamefont{A.~P.} \bibnamefont{Zhuravel}},
  \bibinfo{author}{\bibfnamefont{A.~V.} \bibnamefont{Ustinov}},
  \bibinfo{author}{\bibfnamefont{S.~M.} \bibnamefont{Anlage}},
  \bibnamefont{and} \bibinfo{author}{\bibfnamefont{C.~M.}
  \bibnamefont{Soukoulis}}, \bibinfo{journal}{\prl}
  \textbf{\bibinfo{volume}{107}}, \bibinfo{pages}{043901}
  (\bibinfo{year}{2011}).

\bibitem[{\citenamefont{Ooi et~al.}(2011)\citenamefont{Ooi, Okada, and
  Tanaka}}]{Kazufumi-2011}
\bibinfo{author}{\bibfnamefont{K.}~\bibnamefont{Ooi}},
  \bibinfo{author}{\bibfnamefont{T.}~\bibnamefont{Okada}}, \bibnamefont{and}
  \bibinfo{author}{\bibfnamefont{K.}~\bibnamefont{Tanaka}},
  \bibinfo{journal}{\prb} \textbf{\bibinfo{volume}{84}},
  \bibinfo{pages}{115405} (\bibinfo{year}{2011}).

\bibitem[{\citenamefont{Xu et~al.}(2006)\citenamefont{Xu, Sandhu, Povinelli,
  Shakya, Fan, and Lipson}}]{Xu-2006}
\bibinfo{author}{\bibfnamefont{Q.}~\bibnamefont{Xu}},
  \bibinfo{author}{\bibfnamefont{S.}~\bibnamefont{Sandhu}},
  \bibinfo{author}{\bibfnamefont{M.~L.} \bibnamefont{Povinelli}},
  \bibinfo{author}{\bibfnamefont{J.}~\bibnamefont{Shakya}},
  \bibinfo{author}{\bibfnamefont{S.}~\bibnamefont{Fan}}, \bibnamefont{and}
  \bibinfo{author}{\bibfnamefont{M.}~\bibnamefont{Lipson}},
  \bibinfo{journal}{\prl} \textbf{\bibinfo{volume}{96}},
  \bibinfo{pages}{123901} (\bibinfo{year}{2006}).

\bibitem[{\citenamefont{{Zhang} et~al.}(2008)\citenamefont{{Zhang}, {Genov},
  {Wang}, {Liu}, and {Zhang}}}]{Zhang-2008}
\bibinfo{author}{\bibfnamefont{S.}~\bibnamefont{{Zhang}}},
  \bibinfo{author}{\bibfnamefont{D.~A.} \bibnamefont{{Genov}}},
  \bibinfo{author}{\bibfnamefont{Y.}~\bibnamefont{{Wang}}},
  \bibinfo{author}{\bibfnamefont{M.}~\bibnamefont{{Liu}}}, \bibnamefont{and}
  \bibinfo{author}{\bibfnamefont{X.}~\bibnamefont{{Zhang}}},
  \bibinfo{journal}{\prl} \textbf{\bibinfo{volume}{101}},
  \bibinfo{pages}{047401} (\bibinfo{year}{2008}).

\bibitem[{\citenamefont{Weis et~al.}(2010)\citenamefont{Weis, {Rivi\`ere},
  {Del\'eglise}, {Gavartin}, {Arcizet}, {Schliesser}, and
  {Kippenberg}}}]{Weis-2010}
\bibinfo{author}{\bibfnamefont{S.}~\bibnamefont{Weis}},
  \bibinfo{author}{\bibfnamefont{R.}~\bibnamefont{{Rivi\`ere}}},
  \bibinfo{author}{\bibfnamefont{S.}~\bibnamefont{{Del\'eglise}}},
  \bibinfo{author}{\bibfnamefont{E.}~\bibnamefont{{Gavartin}}},
  \bibinfo{author}{\bibfnamefont{O.}~\bibnamefont{{Arcizet}}},
  \bibinfo{author}{\bibfnamefont{A.}~\bibnamefont{{Schliesser}}},
  \bibnamefont{and} \bibinfo{author}{\bibfnamefont{T.~J.}
  \bibnamefont{{Kippenberg}}}, \bibinfo{journal}{Science}
  \textbf{\bibinfo{volume}{330}}, \bibinfo{pages}{1520} (\bibinfo{year}{2010}).

\bibitem[{\citenamefont{Eichenfield et~al.}(2011)\citenamefont{Eichenfield,
  Winger, Lin, Hill, Chang, and Painter}}]{Eichenfield-2010}
\bibinfo{author}{\bibfnamefont{M.}~\bibnamefont{Eichenfield}},
  \bibinfo{author}{\bibfnamefont{M.}~\bibnamefont{Winger}},
  \bibinfo{author}{\bibfnamefont{Q.}~\bibnamefont{Lin}},
  \bibinfo{author}{\bibfnamefont{J.~T.} \bibnamefont{Hill}},
  \bibinfo{author}{\bibfnamefont{D.~E.} \bibnamefont{Chang}}, \bibnamefont{and}
  \bibinfo{author}{\bibfnamefont{O.}~\bibnamefont{Painter}},
  \bibinfo{journal}{\nat} \textbf{\bibinfo{volume}{472}},
  \bibinfo{pages}{69} (\bibinfo{year}{2011}).

\bibitem[{\citenamefont{Agarwal and Huang}(2010)}]{Agarwal-2010}
\bibinfo{author}{\bibfnamefont{G.~S.} \bibnamefont{Agarwal}} \bibnamefont{and}
  \bibinfo{author}{\bibfnamefont{S.}~\bibnamefont{Huang}},
  \bibinfo{journal}{\pra} \textbf{\bibinfo{volume}{81}},
  \bibinfo{pages}{041803} (\bibinfo{year}{2010}).

\bibitem[{\citenamefont{Kekatpure et~al.}(2010)\citenamefont{Kekatpure,
  Barnard, Cai, and Brongersma}}]{Kekatpure-2010}
\bibinfo{author}{\bibfnamefont{R.~D.} \bibnamefont{Kekatpure}},
  \bibinfo{author}{\bibfnamefont{E.~S.} \bibnamefont{Barnard}},
  \bibinfo{author}{\bibfnamefont{W.}~\bibnamefont{Cai}}, \bibnamefont{and}
  \bibinfo{author}{\bibfnamefont{M.~L.} \bibnamefont{Brongersma}},
  \bibinfo{journal}{\prl} \textbf{\bibinfo{volume}{104}},
  \bibinfo{pages}{243902} (\bibinfo{year}{2010}).

\bibitem[{\citenamefont{Verellen et~al.}(2011)\citenamefont{Verellen, {Van
  Dorpe}, Huang, Lodewijks, Vandenbosch, Lagae, and
  Moshchalkov}}]{Verellen-2011}
\bibinfo{author}{\bibfnamefont{N.}~\bibnamefont{Verellen}},
  \bibinfo{author}{\bibfnamefont{P.}~\bibnamefont{{Van Dorpe}}},
  \bibinfo{author}{\bibfnamefont{C.}~\bibnamefont{Huang}},
  \bibinfo{author}{\bibfnamefont{K.}~\bibnamefont{Lodewijks}},
  \bibinfo{author}{\bibfnamefont{G.}~\bibnamefont{Vandenbosch}},
  \bibinfo{author}{\bibfnamefont{L.}~\bibnamefont{Lagae}}, \bibnamefont{and}
  \bibinfo{author}{\bibfnamefont{V.~V.} \bibnamefont{Moshchalkov}},
  \bibinfo{journal}{Nano Lett.} \textbf{\bibinfo{volume}{11}},
  \bibinfo{pages}{391} (\bibinfo{year}{2011}).

\bibitem[{\citenamefont{{Tassin} et~al.}(2012)\citenamefont{{Tassin},
  {Koschny}, and {Soukoulis}}}]{Tassin-2012}
\bibinfo{author}{\bibfnamefont{P.}~\bibnamefont{{Tassin}}},
  \bibinfo{author}{\bibfnamefont{T.}~\bibnamefont{{Koschny}}},
  \bibnamefont{and} \bibinfo{author}{\bibfnamefont{C.~M.}
  \bibnamefont{{Soukoulis}}}, \bibinfo{journal}{Physica B}
  \textbf{\bibinfo{volume}{407}},
  \bibinfo{pages}{4062} (\bibinfo{year}{2012}).

\bibitem{SupplMat}See Supplemental Material for more results of the radiating two-oscillator model, for a description of the experimental samples and procedures, 
and for a comparison of the reflection and transmission between model and experiment.

\bibitem[{\citenamefont{Akulshin et~al.}(1998)\citenamefont{Akulshin, Barreiro,
  and Lezama}}]{Akulshin-1998}
\bibinfo{author}{\bibfnamefont{A.~M.} \bibnamefont{Akulshin}},
  \bibinfo{author}{\bibfnamefont{S.}~\bibnamefont{Barreiro}}, \bibnamefont{and}
  \bibinfo{author}{\bibfnamefont{A.}~\bibnamefont{Lezama}},
  \bibinfo{journal}{\pra} \textbf{\bibinfo{volume}{57}}, \bibinfo{pages}{2996}
  (\bibinfo{year}{1998}).

\bibitem[{\citenamefont{Taubert et~al.}(2012)\citenamefont{Taubert, Hentschel,
  Kastel, and Giessen}}]{Taubert-2012}
\bibinfo{author}{\bibfnamefont{R.}~\bibnamefont{Taubert}},
  \bibinfo{author}{\bibfnamefont{M.}~\bibnamefont{Hentschel}},
  \bibinfo{author}{\bibfnamefont{J.}~\bibnamefont{Kastel}}, \bibnamefont{and}
  \bibinfo{author}{\bibfnamefont{H.}~\bibnamefont{Giessen}},
  \bibinfo{journal}{Nano Letters} \textbf{\bibinfo{volume}{12}},
  \bibinfo{pages}{1367} (\bibinfo{year}{2012}).

\bibitem[{\citenamefont{Verslegers et~al.}(2012)\citenamefont{Verslegers, Yu,
  Ruan, Catrysse, and Fan}}]{Verslegers-2012}
\bibinfo{author}{\bibfnamefont{L.}~\bibnamefont{Verslegers}},
  \bibinfo{author}{\bibfnamefont{Z.}~\bibnamefont{Yu}},
  \bibinfo{author}{\bibfnamefont{Z.}~\bibnamefont{Ruan}},
  \bibinfo{author}{\bibfnamefont{P.~B.} \bibnamefont{Catrysse}},
  \bibnamefont{and} \bibinfo{author}{\bibfnamefont{S.}~\bibnamefont{Fan}},
  \bibinfo{journal}{\prl} \textbf{\bibinfo{volume}{108}},
  \bibinfo{pages}{083902} (\bibinfo{year}{2012}).

\end{thebibliography}

\end{document}